\documentclass[aps,prl,twocolumn,superscriptaddress]{revtex4-1}

\usepackage{graphicx}

\begin{document}

\title{Highly anisotropic superconducting gap in underdoped Ba$_{1-x}$K$_x$Fe$_2$As$_2$}

\author{H. Kim}
\affiliation{Ames Laboratory and Department of Physics \& Astronomy, Iowa State University, IA 50011, USA}

\author{M. A. Tanatar}
\affiliation{Ames Laboratory and Department of Physics \& Astronomy, Iowa State University, IA 50011, USA}

\author{ Bing Shen}
\affiliation{ Institute of Physics, Chinese Academy of Sciences, Beijing 100190, P. R. China }

\author{Hai-Hu Wen}
\affiliation{ Institute of Physics, Chinese Academy of Sciences, Beijing 100190, P. R. China }
\affiliation{ National Laboratory of Solid State Microstructures and Department of Physics, Nanjing University, Nanjing 210093, P. R. China}

\author{R. Prozorov}
\email[Corresponding author: ]{prozorov@ameslab.gov}
\affiliation{Ames Laboratory and Department of Physics \& Astronomy, Iowa State University, IA 50011, USA}

\date{11 May 2011}

\begin{abstract}
The in-plane London penetration depth, $\Delta\lambda(T)$, was measured using a tunnel diode resonator in single crystals of Ba$_{1-x}$K$_x$Fe$_2$As$_2$ for five doping levels $x$ ranging from underdoped, $T_c$=11~K ($x$=0.17), to optimally doped, $T_c=$38 K ($x_\textmd{\scriptsize opt}$= 0.35). In the optimally doped samples, $\Delta\lambda(T)$ shows exponential saturation in $T \to 0$ limit suggesting a fully gapped superconductivity. The lowest-$T_c$ samples show much stronger non-exponential variation of $\Delta\lambda(T)$. Fitting the data to a power-law, $\Delta\lambda(T)= AT^n$, reveals a monotonic decrease of the exponent $n$ with $x$ towards the underdoped edge of the superconducting dome. Comparison with $n \approx 1.2$ reported in KFe$_2$As$_2$ ($T_c$=3.5 K , $x=$ 1), suggests a dome-like variation of $n$ with $x$, implying an evolution of the topology of the superconducting gap from full and isotropic in the center of the dome towards strongly anisotropic and, eventually, nodal at the dome edges.

\end{abstract}

\pacs{74.25.N, 74.20.Rp,74.70.Xa}

\maketitle


The experimental determination of the symmetry of the superconducting gap is important for understanding the mechanism of superconductivity in iron-based superconductors \cite{Mazin2010Nature,Wang2011}. Measurements of London penetration depth \cite{Gordon2009,Gordon2009a,Martin2010}, thermal conductivity \cite{Tanatar2010,Reid2010} and specific heat \cite{Gofryk2011,Bud'ko2009,Hardy2010} in electron doped Ba(Fe$_{1-x}$Co$_x$)$_2$As$_2$ (BaCo122) suggest that superconducting gap shows strong evolution with doping, developing nodes at the dome edges \cite{Reid2010,Hirschfeld2010physics,Maiti2011}. This doping-evolution is consistent with observations of a fully gapped superconductivity in effectively close to optimally-doped LiFeAs \cite{Borisenko2010,Inosov2010,Kim2011,Tanatar2011} and nodal superconductivity in effectively overdoped KFe$_2$As$_2$  \cite{Fukazawa2009,Dong2010,Hashimoto2010}. It is also consistent with predicted doping-evolution for the $s_{\pm}$ model \cite{Chubukov2009,Maiti2011}. On the other hand, nodal behavior is observed at all doping levels in isovalently substituted BaFe$_2$(As$_{1-x}$P$_x$)$_2$ (BaP122) \cite{Hashimoto2010a}. This noteworthy difference in two systems based on the same parent compound prompts a detailed study of the hole doped Ba$_{1-x}$K$_x$Fe$_2$As$_2$ (BaK122). The superconducting gap in BaK122 was studied intensively using ARPES \cite{Ding2008,Nakayama2011}, NMR \cite{Li2011}, penetration depth \cite{Martin2009,Hashimoto2009} and thermal conductivity \cite{Luo2009,Dong2010}, however, no systematic doping - dependent study reaching the dome edges was undertaken so far.

In this work we study the evolution of the temperature dependence of in-plane London penetration depth, $\Delta\lambda(T)$, in high quality single crystals of Ba$_{1-x}$K$_x$Fe$_2$As$_2$. We find that the optimally doped samples show exponentially weak temperature dependence in $T \to 0$ limit, suggesting a fully gapped superconductivity. This conclusion is consistent with the temperature-dependent superfluid density in these samples, which can be well fitted using self-consistent $\gamma$-model with two full gaps in the clean limit \cite{Kogan2009}. The lowest-$T_c$ samples show an exceptionally strong sub-quadratic temperature dependence. Fitting the experimental $\Delta\lambda(T)$ below $T_c/3$ to a power-law, $\Delta\lambda(T)= AT^n$, we find a monotonic decrease of the  exponent $n$ with concomitant sharp increase of the pre-factor $A$ towards the low $x$ edge of the superconducting dome. Comparison with close to $T$-linear behavior found in heavily overdoped KFe$_2$As$_2$ \cite{Hashimoto2009}, suggests a universal development of nodes at the edges of the superconducting dome in both electron- and hole-doped BaFe$_2$As$_2$ - based superconductors.

Single crystals of Ba$_{1-x}$K$_x$Fe$_2$As$_2$ were grown using high temperature FeAs flux method \cite{Luo2008}. $\Delta\lambda(T)$ was measured using tunnel-diode resonator technique \cite{Degrift1975,Prozorov2006}. Placing a sample into the inductor causes the shift of the resonant frequency, $\Delta f(T)=-G4\pi\chi(T)$. Here $4\pi\chi(T)$ is magnetic susceptibility and $G$ is a calibration constant determined by physically pulling the sample out of the coil. With the characteristic sample size, $R$, $4\pi\chi=(\lambda/R)\tanh (R/\lambda)-1$, from which $\Delta \lambda$ can be obtained \cite{Prozorov2000,Prozorov2006}. The excitation field in the inductor, $H_{ac} \sim 20$ mOe, is much smaller than $H_{c1}$.

\begin{figure}
\includegraphics[width=1.0\linewidth]{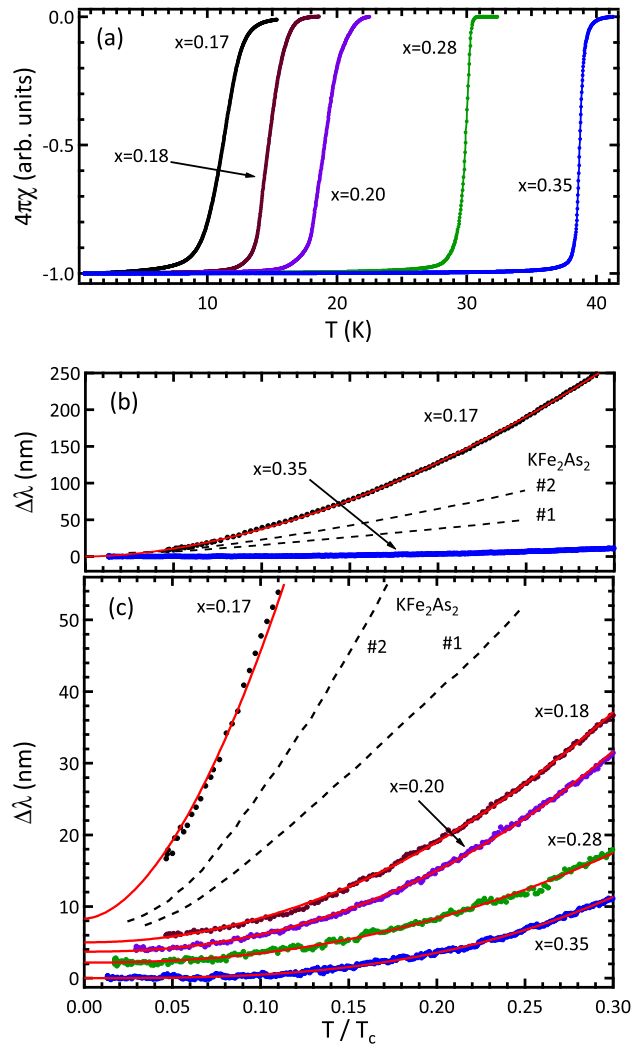}
\caption{\label{fig1} (a) Normalized $\Delta\lambda(T)$. (b) and (c) Low-temperature parts of $\Delta\lambda(T)$. Experimental data are displayed with solid circles. Solid red line represent power-law fitting curves. Two dashed lines are data for pure KFe$_2$As$_2$ from Ref.~\onlinecite{Hashimoto2010}.}
\end{figure}

To compare sharpness of the superconducting transition, Fig.~\ref{fig1}(a) shows normalized RF susceptibility of Ba$_{1-x}$K$_x$Fe$_2$As$_2$ samples used in this study. The superconducting transition remains quite sharp even for the most underdoped samples where $T_c(x)$ is very sensitive to small variations of $x$. The values of $x$ were determined by the empirical fit \cite{Tanatar2011BaKAniso} of the experimental $T_c (x)$ data \cite{Luo2008,Avci2011}. The values of $T_c$ were determined from the position of the maximum in the first derivative, $d\Delta\lambda(T)/dT$. In our samples we obtained $T_c$ of 11.2, 14.5, 18.6, 30.0, and 38.7 K, corresponding to potassium concentrations of $x=$0.17, 0.18, 0.20 (all $\pm$0.01), 0.28 ($\pm$0.02), and 0.35 ($\pm$0.03), respectively. The low-temperature variation of $\Delta\lambda(T)$ up to $T_c/3$ is shown in Figs.~\ref{fig1}(b) and (c). Figure~ \ref{fig1}(b) compares the data for limiting compositions $x$=0.17 and 0.35, revealing a big difference in the magnitude of $\Delta\lambda(T)$ . Two curves for pure KFe$_2$As$_2$ are shown for reference \cite{Hashimoto2010}.  Figure \ref{fig1}(c) shows $\Delta\lambda(T)$ on the same scale for all concentrations. The data are offset for clarity and red lines represent the power-law fit. However, a closer look shows significant deviations of the data in heavily underdoped ($x=0.17$) and in optimally doped ($x=0.35$) samples. At lowest temperatures, it becomes significantly sub-quadratic for the former and closer to exponential for the latter.

First we attempted to fit the data for two highest $T_c$ samples to the single gap $s$-wave BCS function, $\Delta \lambda(T)/\lambda(0)=\sqrt{\pi \Delta_0/2k_BT}\exp(-\Delta_0/k_BT)$, where $\Delta_0$ is the size of gap at $T=0$. The $\Delta_0$ values from the best fittings are $0.73~k_BT_c$ and $0.87~k_BT_c$ for $x=0.28$, and 0.35, respectively. While the fit quality was good, both $\Delta_0$ values are much smaller than in single full-gap superconductors where $\Delta_0 = 1.76~k_BT_c$. Such small gaps are expected in superconductors with $\Delta_\textmd{\scriptsize min} < \Delta_\textmd{\scriptsize max}$, either due to gap angular variation (anisotropy) or variation between different Fermi surface sheets.

A standard way to analyze $\Delta\lambda(T)$ is to fit it from the lowest temperature up to $T_\textmd{\scriptsize up} \approx T_c/3$. In a single -gap $s$-wave superconductor this limit is determined by reaching nearly constant value of the superconducting gap $\Delta_0$, below which the temperature dependence is exponential. For various nodal gaps, the dependence is expected to be power-law, $T$-linear for line nodes and $T^2$ for point nodes  in clean limits. For the anisotropic gap or multi-gap superconductors with the variation of the gap magnitude over the Fermi surface between $\Delta_\textmd{\scriptsize max}$ and $\Delta_\textmd{\scriptsize min}$, the $ T_\textmd{\scriptsize up}$ is determined by $\Delta_\textmd{\scriptsize min}$, while $T_c$ by $\Delta _\textmd{\scriptsize max}$, so that $T_\textmd{\scriptsize up}$ range of characteristic temperature dependence can be smaller than $T_c/3$.

\begin{figure}
\includegraphics[width=1.0\linewidth]{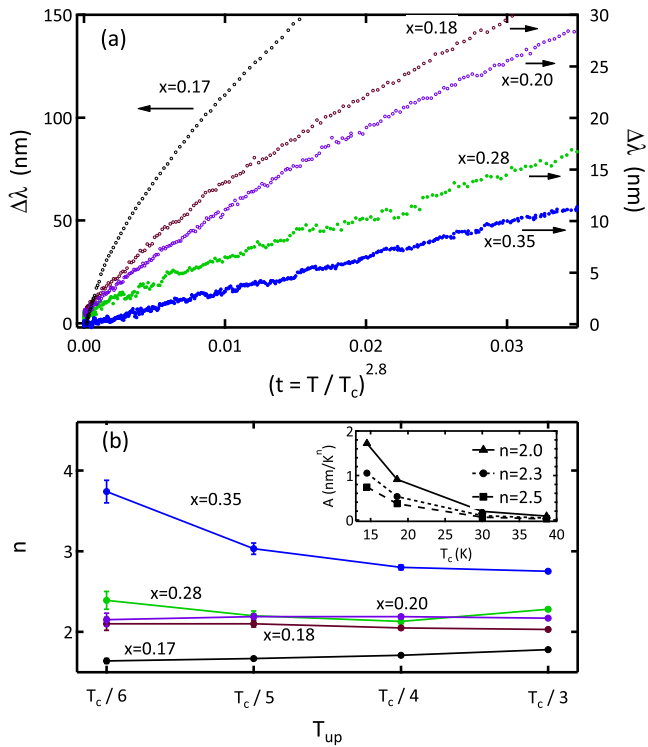}
\caption{\label{fig2}(a) $\Delta\lambda$ vs. $(T/T_c)^{2.8}$ up to $T_c/3$. (b) Variation of exponent $n$ obtained from a power-law fitting, $\Delta\lambda =A  (T/T_c)^n$, as a function of the upper end temperature of the fitting range. Inset: $A$ vs. $T_c$. The three curves represented by triangle, circle, and square were obtained with fixed $n=$2.0, 2.3, and 2.5, respectively.}
\end{figure}

We therefore check the alteration of the fitting parameters by choosing different temperatures for the upper limit, $ T_\textmd{\scriptsize up} <T_c/3$. The dependence of $n$ and $A$ on $ T_\textmd{\scriptsize up}$ is shown in Fig.~\ref{fig2}(b). The highest-and lowest-$T_c$ samples exhibit monotonic increase and decrease of $n$ on $T_\textmd{\scriptsize up} \to 0$, approaching very different values of 4 and 1.5, respectively.

The exponents $n$ for samples with $x=0.18$, 0.20, and 0.28 do not show a significant variation with $T_\textmd{\scriptsize up}<T_c/3$, indicating robust power-law behavior, but show systematic increase of $n$ with $x$. The decrease of $n$ with decrease of $x$ can be clearly seen by in the top panel of Fig.~\ref{fig2}, in which all data are plotted vs $(T/T_c)^{2.78}$, where $n=2.78$ is the exponent for the optimally doped samples. The dependence of the power-law pre-factor $A$ on $T_c$ was analyzed by fixing $n$=2.0, 2.3, and 2.5 and is shown in the inset revealing a significant increase with decreasing $T_c$. The $A$ value for sample $x$=0.17 is 30 nm/K$^{1.78}$, out of scale for the plot and is not shown.

\begin{figure}
\includegraphics[width=1.0\linewidth]{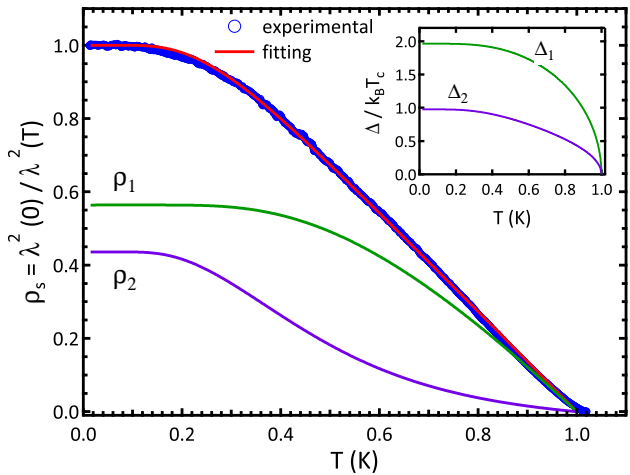}
\caption{\label{fig3}Symbols: superfluid density $\rho_s(T)=\lambda^2(0)/\lambda^2(T)$ calculated with $\lambda(0)=200$ nm in Ba$_{0.65}$K$_{0.35}$Fe$_2$As$_2$ \cite{Li2008}. Solid lines represent the fit to a two-gap $\gamma$ model, $\rho_s=\gamma\rho_1 +(1-\gamma)\rho_2$. Inset: superconducting gaps $\Delta_1(T)$ and $\Delta_1(T)$ calculated self-consistently during the fitting.
}
\end{figure}

A smaller than weak - coupling value of $\Delta_{min}$ obtained from low-temperature BCS formula implies two - gap superconductivity and the analysis must be extended to the full - temperature range. The most convenient quantity is the superfluid density, which can be calculated from the first principles. In the optimally doped samples, we fit the data using clean-limit $\gamma-$model \cite{Kogan2009}. Symbols in Fig.~\ref{fig3} show superfluid density,  $\rho_s(T)=\lambda^2(0)/\lambda^2(T)$, for the sample with $x=0.35$ calculated from $\lambda(T) = \Delta \lambda(T)+\lambda(0)$ with $\lambda(0)=200$ nm \cite{Li2008}. Solid lines show self-consistent $\gamma$-model fit for two-full-gap superconducting state \cite{Kogan2009} with $\rho_s = \gamma \rho_1 + (1-\gamma)\rho_2$, where $\rho_1$ and $\rho_2$ are partial superfluid densities. Insert shows two superconducting gaps $\Delta_1$ and $\Delta_2$ calculated during the fitting procedure. The estimated gap values are 6.5 and 3.3 meV. Specific heat jump produced the value of $\sim 6$ meV for the larger gap \cite{Mu2009Cp}.

Upon departure from optimal doping, the exponent $n$ shows notable evolution with $x$ decreasing from 4 to about 1.5. London penetration depth is very sensitive to pair-breaking disorder, modifying $\Delta\lambda(T)$ at low temperatures \cite{Bang2009,Kogan2009}. In Ba122 - derived compounds it was also suggested experimentally
\cite{Hashimoto2009,Martin2009,Kim2010a,Gordon2010}. Within the $s_\pm$ theory \cite{Mazin2008}, $\lambda(T)$ should be exponential in the clean limit \cite{Kim2011,Imai2011}. However, pair-breaking scattering (which in this case can be caused by non-magnetic impurities and dopant ions) turn the behavior into a power-law with the exponent $n$ approaching 2 in the dirty limit \cite{Vorontsov2009,Bang2009,Glatz2010}. Since the superconductivity in BaK122 is induced by doping, we cannot ignore the effect of disorder on the variation of exponents. However, it would be natural to expect increase of scattering with $x$, and thus decrease of the exponent, opposite to the trend in our data. Similarly, disorder effect cannot explain nodal state in the end member of BaK122, very pure KFe$_2$As$_2$ with $n$=1.2 \cite{Hashimoto2010,Dong2010}. In addition, our most underdoped sample shows the exponent $n=1.5$ clearly well below the limiting value of 2 for pair-breaking scattering. Thus we conclude that the variation of the exponent $n$, found in our study, is caused by the changes in the superconducting gap structure with doping.

\begin{figure}
\includegraphics[width=1.0\linewidth]{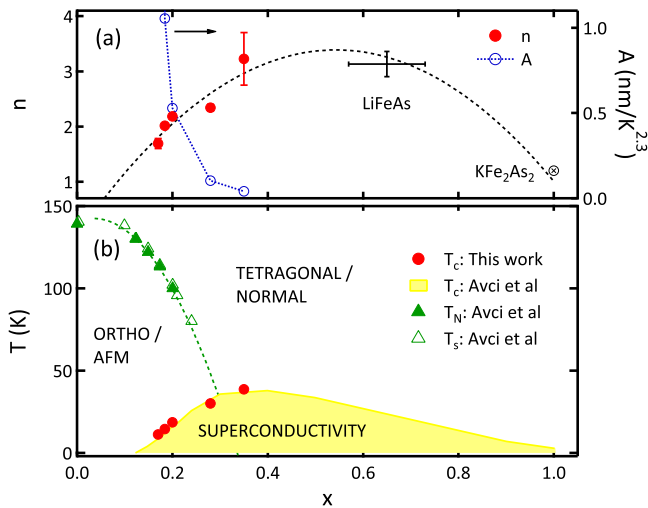}%
\caption{\label{fig4}(a) $n$ vs. $x$ diagram. The red solid circles represent $n$ from various temperature range between $T_c/6$ and $T_c/3$. The exponent $n\approx1$ and 3.1 in KFe$_2$As$_2$ \cite{Hashimoto2010} and LiFeAs \cite{Kim2011}, respectively, are from elsewhere. (b) $T$-$x$ phase diagram. The AFM region is from neutron scattering data \cite{Avci2011}.}
\end{figure}

The evolution of the power-law behavior in Ba$_{1-x}$K$_x$Fe$_2$As$_2$ superconductors is summarized in Fig.~\ref{fig4}(a). Solid circles show exponent $n$ with the error bars estimated from the fitting to the different temperature ranges (such as shown in Fig.~\ref{fig2}(b)) and open circles show the pre-factor $A$ calculated for a fixed exponent $n=2.3$. Also shown are the exponents for two stoichiometric (clean) compounds, KFe$_2$As$_2$ \cite{Hashimoto2010} and LiFeAs \cite{Kim2011}. The dashed line represents our picture of the exponent variation with doping that, in our opinion, reflects developing anisotropy of the superconducting gap. To relate to the phase diagram, Fig.~\ref{fig4}(b) shows magnetic and superconducting transitions vs doping from neutron scattering \cite{Avci2011}. The pre-factors of the power-law fit show a sharp increase in the antiferromagnetic (AFM) region, similar to FeCo122 \cite{Gordon2010a}, which indicates microscopic coexistence of superconductivity and long-range magnetic order \cite{Fernandes2010}.

The value of the exponent $n$=1.5 in the sample with $x$=0.17 suggests the existence of line nodes in the superconducting gap in heavily underdoped compositions. This value is too small to be explained in the full-gap $s_\pm$ scenario with strong pairbreaking scattering \cite{Kim2010a}. In superconductors with line nodes the exponent $n$ varies between $n$=1 in the clean limit and $n$=2 in the dirty limit \cite{Hirschfeld1993dwave}. Our values of $n$ are well within this range and outside the dirty $s_\pm$.

We would like to emphasize that in electron-doped Ba122 a clear signature of nodes appears only for the $c$-axis as shown by the penetration depth \cite{Martin2010} and thermal conductivity measurements \cite{Reid2010}. Here we find that in the hole-doped Ba122 the nodes seem to appear even in the $ab$-plane measurements.
This may be indicative of the different nodal structure in these two systems. In both cases, the nodes develop far from the optimal doping.

Strong dependence of the superconducting gap structure on the shape of the three-dimensional Fermi surface is predicted in the $s_\pm$ model \cite{Hirschfeld2010physics}. In the same model, the gap also evolves with doping due to changing nesting conditions, in rough agreement with the evolution we see here \cite{Chubukov2009}. However, the nodal superconducting gap can also be explained by the phase transition from fully-gapped $s_\pm$ to nodal $d$-wave, which have relatively small difference in ground state energy \cite{Chubukov2009}. However, this model does not explain a gradual evolution of the exponent $n$ with doping.

In conclusion, the measurements of $\lambda(T)$ in Ba$_{1-x}$K$_x$Fe$_2$As$_2$ suggest doping-dependent evolution of the superconducting gap from full and isotropic at optimal doping to highly anisotropic and, eventually, nodal in heavily underdoped samples.

We thank A. Chubukov, P. Hirschfeld and L. Taillefer for useful discussions. The work at Ames was supported by the U.S. Department of Energy, Office of Basic Energy Sciences, Division of Materials Sciences and Engineering under contract No. DE-AC02-07CH11358. Work in China was supported by the Ministry of Science and Technology of China, project 2011CBA0102.

\end{document}